\documentstyle[12pt]{article}
\input epsf
\begin{document}
\vskip 0.5cm
\centerline{\large Mass Differences within Isotopic Multiplets in a SUSY} 
\centerline{\large Electro-weak Theory}
\vskip 0.2cm
%
% authors
\begin{center}
Guanghua Xu$^{\ast}$\\
{\sl University of California, Riverside, California 92521}\\
\end{center}
%
%  Abstract
\begin{abstract}
Based on the idea that electromagnetism is responsible for the mass differences 
within isotopic multiplets, and possibly also for the whole mass of the 
electron, a supersymmetric gauge theoretical model based on the group 
$SU(2)_{L} \times SU(2)_{R} \times U(1)_{Y}$ is constructed. Under
some reasonable assumptions to the SUSY particle spectrum, a correct sign for
the mass difference within an isotopic multiplet is 
obtained. This might provide a possible scenario to understand the old puzzle 
of the proton-neutron mass difference.      
\end{abstract}
\noindent{$Introduction -$} It had challenged and frustrated generations of
physicists 
to apply the idea that electromagnetic and weak interactions are responsible 
for mass differences within isotopic multiplets, e.g. $\Delta m|_{d-u}$, and 
possibly also for the whole mass of the electron to calculations for they 
always gave a wrong sign[1,2]. In the last few decades, as physicists 
understand more the interactions and fundamental structures of matters, 
they tend to believe that[1] isotopic symmetry is not a fundamental 
symmetry in strong interaction, and the false impression is due to the small 
u-d quark mass difference, though it is comparable to the quark masses 
themselves (about a few Mev), on the typical strong-interaction scale (about 
200-400 Mev). This sort of view may well be correct, but a rigorous 
experimental proof  will not be easy[1]. For many reasons, the idea that 
$\Delta m|_{d-u}$ is due to electromagnetic and weak interactions is still very 
attractive, although there are difficulties in 
calculations of some physical quantities. Alternatively, it is natural to ask 
ourselves whether the previous incorrect results in the calculations are due to 
the limit of our theoretical understanding of the nature? For this reason, the 
author and his collaborator once considered a supersymmetric extension of an 
$SU(2) \times U(1)$ toy model[3] and nicely obtained a correct sign for the 
mass difference within an isodoublet. Although it is just a toy model, the 
result is still very encouraging. The question is whether we can construct a 
more realistic model which should be consistent with the Standard Model. 

    In this letter, I will study a supersymmetric extension of an 
$SU(2)_{L} \times SU(2)_{R} \times U(1)_{Y}$ model, which was originally 
suggested by S. Weinberg[2], and will discuss the mass difference within an 
isodoublet in this model. It is supposed that the weak and electromagnetic 
gauge group $SU(2)_{L} \times U(1)_{Y}$ is part of a larger gauge group 
$SU(2)_{L} \times SU(2)_{R} \times U(1)_{Y}$. We don't see effects of the 
gauge bosons associated with such transformations, so we must suppose that 
they are very heavy[1]. Fortunately these vector bosons can be almost 
arbitrarily heavy, and still produce the necessary mass shifts.  The mass 
difference within an isotopic multiplet in this model is due to the "type 1" 
mass relation[2], which guarantees that the mass difference does not arise 
from graphs involving virtual scalar bosons, whose properties are 
almost entirely unknown.  

\noindent{$The$
$Supersymmetric$ $SU(2)_{L} \times SU(2)_{R} \times U(1)_{Y}$ $Model - $}
The supersymmetric generalization[4] of this model consists of the fields 
listed in Table 1. Comparing to the original gauge field model[2], here 
$\Phi_{2}$ 
is added to take care of the problem in $\beta$ decay (see Freedman's paper in 
[2]), and ${\Phi}'_{1}$ and ${\Phi}'_{2}$ are needed for generating masses for 
the supersymmetric partners of the gauge particles, and $N_{i}$'s are 
responsible for the existence of a unique ground state which breaks 
$SU(2)_{L} \times SU(2)_{R} \times U(1)_{Y}$ to $U(1)_{EM}$ at tree level for 
this unbroken supersymmetric model. 

    The scalar potential $V$ in the SUSY $SU(2)_{L} \times SU(2)_{R} \times 
U(1)_{Y}$ model is  
%eq1
\begin{eqnarray} 
  V = \frac{1}{2} [D^{a}_{L} D^{a}_{L} + D^{a}_{R} D^{a}_{R} + (D')^{2}] + 
      F_{i}^{*} F_{i} ,
\end{eqnarray}
where 
%eq2
\begin{eqnarray}
  D^{a}_{L(R)} = \frac{g_{L(R)}}{2} A_{iL(R)}^{*} {\tau}^{a}_{ij} A_{jL(R)}, 
  \ \ \ D' = \frac{g_{Y} y_{i}}{2} A_{i}^{*} A_{i} , \ \ \ F_{i} = \frac{ 
  \partial W}{\partial A_{i}} ,
\end{eqnarray} 
with $A_{iL(R)}$ as scalar fields transforming as doublets in $SU(2)_{L(R)}$ 
respectively, $A_{i}$ as scalar fields listed in Table 1, and  $W =  
(h_{m} {\epsilon}_{ij} {\Phi}_{1}^{i} {\Phi}_{1}^{\prime j} + l_{m} {\epsilon}_{ij} 
{\Phi}_{2}^{i} {\Phi}_{2}^{\prime j} + k_{m} H_{ij} H_{ji} + s_{m}) 
N_{m}$$+f {\epsilon}_{ij} F_{R}^{i} H^{jk} F_{L}^{k}$, where $m = 1, 2, 3$.  

    From eqs.1 and 2, the nonzero vacuum expectation values of the scalar 
fields are 
%eq3
\begin{eqnarray}
  H & = & \frac{1}{\sqrt{2}} \left( \begin{array}{cc}  v {\ }\  & {\ } 0 \\ 
  0 {\ } & {\ }\  v \end{array}   \right), \ \  \Phi_{1} = \frac{1}{\sqrt{2}} 
  \left( \begin{array}{c} 0 \\ v_{1}  \end{array}  \right),  \ \    
  {\Phi}'_{1} = \frac{1}{\sqrt{2}} \left( \begin{array}{c} v_{1} \\
  0 \end{array}  \right) , \nonumber \\ 
  \Phi_{2} & = & \frac{1}{\sqrt{2}} \left( \begin{array}{c} 0 \\ v_{2} 
  \end{array}  \right), \ \ {\Phi}'_{2} = \frac{1}{\sqrt{2}} \left( 
  \begin{array}{c} v_{2} \\ 0 \end{array}  \right) ,
\end{eqnarray}
which break $SU(2)_{L} \times SU(2)_{R} \times U(1)_{Y}$ down to $U(1)_{EM}$. 
The constants $v$, $v_{1}$ and $v_{2}$ are related to $h_{m}$, $l_{m}$, 
$k_{m}$, $s_{m}$ of eq.2 by $\frac{1}{2} h_{m} v_{1}^{2} + \frac{1}{2} l_{m} 
v_{2}^{2} + k_{m} v^{2} + s_{m} = 0$. As can be seen from eqs.1, 2 and 3, the 
scalar potential has $V_{min} = 0$, thus implying that the theory remains 
supersymmetric. 

    By considering the interaction terms after spontaneous gauge symmetry 
breaking, we can have the following mass eigenstates,
%eq4 
\begin{eqnarray}
  &   & \left( \begin{array}{c} A^{\mu} ({\lambda}_{A}) \\ Z^{\mu}_{1} ({
  \lambda}_{Z_{1}}) \\ Z^{\mu}_{2} ({\lambda}_{Z_{2}}) \end{array} \right) = 
  \left( \begin{array}{ccc} a_{11} {\ } & a_{12} {\ } & a_{13}  \\ a_{21} {\ } 
  & a_{22} {\ } & a_{23}  \\ a_{31} {\ } & a_{32} {\ } & a_{33}  \end{array}   
  \right)  \left( \begin{array}{c} A^{\mu L}_{3} ({\lambda}_{L}^{3}) \\ 
  A^{\mu R}_{3} ({\lambda}_{R}^{3}) \\ B^{\mu} ({\lambda}_{Y}) \end{array} 
  \right),  \nonumber \\
  &   & {\zeta}_{A} = {\zeta}_{1} = \left( \begin{array}{c} -i {\lambda}_{A} 
  \\ i {\bar{{\lambda}}}_{A} \end{array}\right) ,  \ \   
  {\zeta}_{Z_{1,2}} = {\zeta}_{2,3} = \left( \begin{array}{c} -i 
  {\lambda}_{2,3} \\ i {\bar{{\lambda}}}_{Z_{1,2}} \end{array}\right) ,
\end{eqnarray} 
with 
%eq5
\begin{eqnarray}   
  -i {\lambda}_{i} & = & \frac{1}{2 m_{Z_{i}}} [ {v_{1}} (g_{L} a_{i1} 
  - g_{Y} a_{i3})({\tilde{\Phi}}_{1}^{\prime 1} - {\tilde{\Phi}}_{1}^{2} ) + {v
  } (g_{L} a_{i1} - g_{R} a_{i2})({\tilde{H}}_{11} - {\tilde{H}}_{22}) + 
  \nonumber \\
  &   & {v_{2}} (g_{L} a_{i2} - g_{Y} a_{i3}) ( {\tilde{\Phi}}_{2}^{\prime 1} - 
  {\tilde{\Phi}}_{2}^{2} ) ], \nonumber \\   
  i & = & 2,3, \ \ \ m_{A} = m_{{\zeta}_{A}} = 0, \nonumber \\ 
  {{m^{2}}_{Z_{1,2}}} & = & {{m^{2}}_{{\zeta}_{Z_{1,2}}}} =  \{ [
  g^{2}_{L} ( v^{2} + v_{1}^{2} ) + g^{2}_{R} (v^{2} + v_{2}^{2} )+ g^{2}_{Y}
  (v_{1}^{2} + v_{2}^{2} )] \pm ( [g^{2}_{L} (v^{2} + v_{1}^{2} ) + 
  \nonumber \\ 
  &   & \ \ \ \ \ \ \ \ \ \ \ \ \ \ \ g^{2}_{R}(v^{2} + v_{2}^{2} ) + g^{2}_{Y
  } (v_{1}^{2} + v_{2}^{2} )]^{2} - 4 (v^{2} v_{1}^{2} + v^{2} v_{2}^{2} + 
  v_{1}^{2} v_{2}^{2}) (g^{2}_{L} g^{2}_{R} + \nonumber \\ 
  &   & \ \ \ \ \ \ \ \ \ \ \ \ \ \ \ (g^{2}_{L} g^{2}_{R} + g^{2}_{L} g^{2
  }_{Y} + g^{2}_{R} g^{2}_{Y}))^{2} \} /8 ,
\end{eqnarray}
where the $(a_{ij})$ is orthogonality matrices. Their elements can be 
determined by orthogonality condition and the lagrangian.[5]

    The Lagrangian of the SUSY model can be written as 
%eq6
\begin{eqnarray}
  {\cal L}_{int} & = & \bar{\psi} i {\gamma}^{\mu} ( g_{L} {\vec{\tau}} 
  \cdot {\vec{A_{\mu}}}^{L} P_{L} + g_{R} {\vec{\tau}} \cdot {\vec{A_{\mu}}
  }^{R} P_{R} + g_{Y} y_{L} B_{\mu} ) \psi / 2+ g_{L} {\bar{\tilde{\omega}}}_{L
  }^{-} P_{L} \psi_{2} {\tilde{\psi}}_{1L}^{*} + \nonumber \\ 
  &   & g_{L} {\bar{\psi}}_{1} P_{R
  } {\tilde{\omega}}_{L}^{+} {\tilde{\psi}}_{2L} + g_{L} {\bar{\tilde{\omega
  }}}_{L}^{+} P_{L} \psi_{1} {\tilde{\psi}}_{2L}^{*} + 
  g_{L} {\bar{\psi}}_{2} P_{R} {\tilde{\omega}}_{L}^{-} {\tilde{\psi}
  }_{1L} - g_{R} {\bar{\psi}}_{1} P_{L} {\tilde{\omega}}_{R}^{-c} {\tilde{
  \psi}}_{2R} - \nonumber \\ 
  &   & g_{R} {\bar{\tilde{\omega}}}_{R}^{+c} P_{R} {\psi}_{2} {\tilde
  {\psi}}_{1R}^{*} - g_{R} {\bar{\psi}}_{2} P_{L} {\tilde{\omega}}_{R}^{+c} {
  \tilde{\psi}}_{1R} - g_{R} {\bar{\tilde{\omega}}}_{R}^{-c} P_{R} {\psi}_{1}{
  \tilde{\psi}}_{2R}^{*} + \nonumber \\
  &   & \{ (g_{L} a_{j1} + g_{Y} y_{L} a_{j3}) [ {\bar{\zeta}
  }_{j} P_{L} \psi_{1} {\tilde{\psi}}_{1L}^{*} + {\bar{\psi}}_{1} P_{R} {\zeta
  }_{j} {\tilde{\psi}}_{1L} ] - \nonumber \\ 
  &   & (g_{L} a_{j1} - g_{Y} y_{L}
  a_{j3}) [ {\bar{\zeta}}_{j} P_{L} \psi_{2} {\tilde{\psi}}_{2L}^{*} + {\bar{
  \psi}}_{2} P_{R} {\zeta}_{j} {\tilde{\psi}}_{2L} ] + \nonumber \\
  &   & (g_{R} a_{j2} - g_{Y} y_{L} a_{j3}) [ {\bar{\psi}}_{2} P_{L} {\zeta
  }_{j}^{c} {\tilde{\psi}}_{2R} + {\bar{\zeta}}_{j}^{c} P_{R} {\psi}_{2} {
  \tilde{\psi}}_{2R}^{*}] - \nonumber \\ 
  &   & (g_{R} a_{j2} + g_{Y} y_{L} a_{j3}) [ {\bar{
  \psi_{1}}} P_{L} {\zeta}_{j}^{c} {\tilde{\psi}}_{1R} + {\bar{\zeta}}_{j}^{c} 
  P_{R} {\psi}_{1} {\tilde{\psi}}_{1R}^{*} ] \} / {\sqrt{2}} ,
\end{eqnarray} 
where ${\tilde{\omega}}_{L}^{-}$, ${\tilde{\omega}}_{L}^{+}$, 
${\tilde{\omega}}_{R}^{-}$, ${\tilde{\omega}}_{R}^{+}$ are the SUSY partners 
of the gauge fields $A_{L}^{-}$, $A_{L}^{+}$, $A_{R}^{-}$, $A_{R}^{+}$, 
respectively. 

    Working in $U$ gauge, from eq.6, we can obtain the second order 
$\Delta m|_{d-u}$ in the supersymmetric 
$SU(2)_{L} \times SU(2)_{R} \times U(1)_{Y}$ model as
%eq7
\begin{eqnarray}
  \Delta m|_{d-u} = - \frac{\alpha m}{2 \pi} {\int}_{0}^{1} dx [ d_{1}
  ln(1+ \frac{1-x}{x^{2}} \frac{m_{Z_{1}}^{2}}{m^{2}}) + d_{2} ln(1+ \frac{
  1-x}{x^{2}} \frac{m_{Z_{2}}^{2}}{m^{2}}) ] ,
\end{eqnarray}
where $m$ is the zeroth-order mass of the isodoublet appearing in the
Lagrangian, $m_{Z_{1}}$, $m_{Z_{2}}$ are given by eq.5 and $\alpha = 
{e^{2}}/{4 \pi} = {1}/{137.04}$ with $e^{2}$, $d_{1}$, $d_{2}$ defined by
%eq8
\begin{eqnarray}
  e & = & \frac{g_{L} g_{R} g_{Y}}{({g_{L}}^{2} {g_{R}}^{2} + {g_{L}}^{2} {
  g_{Y}}^{2} + {g_{R}}^{2} {g_{Y}}^{2})^{1/2}}, \nonumber \\  
  d_{1} & = & - \frac{a_{23} (g_{L} a_{21} + g_{R} a_{22})}{a_{13}
  (g_{L} a_{11} + g_{R} a_{12})}, \ \ \ d_{2} = - \frac{a_{33} (g_{L} a_{31} +
  g_{R} a_{32})}{a_{13} (g_{L} a_{11} + g_{R} a_{12})} ,
\end{eqnarray}
where $e^{2}$ appears in eq.7 as the coefficient of the photon term. In view 
of the orthogonality conditions for $a_{ij}$, $d_{1,2}$ satisfy
$d_{1} + d_{2} = 1$.

    Comparing to the result from the pure gauge field model[2], 
%eq9     
\begin{eqnarray}
  \Delta m_{G.F.} |_{d-u} = - \frac{\alpha m}{2 \pi} {\int}_{0}^{1} dx (1+x)[ 
  d_{1} ln(1+ \frac{1-x}{x^{2}} \frac{m_{Z_{1}}^{2}}{m^{2}}) + d_{2} ln(1+ 
  \frac{1-x}{x^{2}} \frac{m_{Z_{2}}^{2}}{m^{2}}) ] ,
\end{eqnarray}
we see that $\Delta m|_{d-u}$ is still negative although it is less negative 
than the result obtaining from the corresponding pure gauge field model. But 
the encouraging thing is that the contribution to $\Delta m|_{d-u}$ from the 
SUSY partners could be positive. This raises some hope for getting a right 
sign for $\Delta m|_{d-u}$.

    As we know, if supersymmetry is really a theory describing the nature, it 
should be broken for no supersymmetry exhibiting in the low energy particle 
spectrum. Therefore, a calculation of $\Delta m|_{d-u}$ from the 
supersymmetric Lagrangian 
with spontaneous gauge symmetry breaking is not complete. We should also 
consider the contribution to $\Delta m|_{d-u}$ due to supersymmetry breaking. 

\noindent{$Supersymmetry$
$Breaking$ $in$ $the$ $SU(2)_{L}$ $\times$ $SU(2)_{R}$ $\times$ 
$U(1)_{Y}$ $Model - $} I will consider explicit soft-supersymmetry breaking in
this section. When SUSY is softly broken, the possible mass terms in two 
component notations are, 
%eq10
\begin{eqnarray}
  &   & {\tilde{\psi}}_{iL}^{*} L^{2}_{i} {\tilde{m}}^{2} {\tilde{\psi}}_{iL}
  + {\tilde{\psi}}_{iR}^{*} R^{2}_{i} {\tilde{m}}^{2} {\tilde{\psi}}_{iR} +
  2 A_{i} {\tilde{m}} m Re {\tilde{\psi}}_{iL}^{*} {\tilde{\psi}}_{iR} -
  {\mu}_{1} {\epsilon}^{\alpha \beta} {\tilde{\Phi}}_{1}^{\alpha}
  {\tilde{\Phi}}_{1}^{\prime \beta} - {\mu}_{2} {\tilde{H}}_{ij} {\tilde{H}}_{ji} -
  \nonumber \\   
  &   & {\mu}_{3} {\epsilon}^{\alpha \beta} {\tilde{\Phi}}_{2}^{\alpha}
  {\tilde{\Phi}}_{2}^{\prime \beta} + (M_{1} /2) {\lambda}_{L}^{a} {\lambda}_{L}^{a} 
  + ({M_{2}}/{2}) {\lambda}_{R}^{a} {\lambda}_{R}^{a} + ({M_{3}}/2) {\lambda}' 
  {\lambda}' .
\end{eqnarray}
This will lead to the mixings among different particles. In principle, it is 
better to obtain mass eigenstates and their corresponding masses numerically. 
For simplicity and analyticity, an analytically worked example is the case 
where ${\mu}_{1} = {\mu}_{2} = {\mu}_{3} =0$. If I further set 
$v'_{1} = v^{\prime \prime}_{1} = v_{1}$, $v'_{2} = v''_{2} = v_{2}$, 
$M_{1} = M_{2} = M_{3} = M_{0}$, then the mass eigenstates and their masses in 
the SUSY breaking model will be given in the following. Note that the 
superpotential in this case is not necessary to have the form specified in 
eq.1 and it will allow a more general set of vacuum expectation values than 
that of eq.3, e.g.
%eq11 
\begin{eqnarray}
  {\Phi}_{1} = \frac{v'_{1}}{\sqrt{2}} \left( \begin{array}{c} 0 \\ 1
  \end{array} \right), \ \ {\Phi}'_{1} = \frac{v^{\prime \prime}_{1}}{\sqrt{2}}
  \left( \begin{array}{c} 1 \\ 0 \end{array} \right), \ \ 
  {\Phi}_{2} = \frac{v'_{2}}{\sqrt{2}} \left( \begin{array}{c} 0 \\ 1
  \end{array} \right), \ \ {\Phi}'_{2} = \frac{v^{\prime \prime}_{2}}{\sqrt{2}}
  \left( \begin{array}{c} 1 \\ 0 \end{array} \right) .
\end{eqnarray}

\noindent{1. Mixing of scalar-quarks:}

    The mass eigenstates and their masses are 
%eq12
\begin{eqnarray}
  {\tilde{\psi}}_{iI} & = & {\tilde{\psi}}_{iL} cos {\theta}_{i} + 
  {\tilde{\psi}}_{iR} sin {\theta}_{i}, \ \ \ 
  {\tilde{\psi}}_{iII} = - {\tilde{\psi}}_{iL} sin {\theta}_{i} + 
  {\tilde{\psi}}_{iR} cos {\theta}_{i}, \nonumber \\ 
  tan 2 {\theta}_{i} & = & {2 A_{i} m}/[(L_{i}^{2} - R_{i}^{2}) {\tilde{m}
  }^{2}],  \nonumber \\ 
  {\mu}_{iI,II}^{2} & = & m^{2} + \{ (L_{i}^{2} + R_{i}^{2}) {\tilde{m}}^{2} 
  \pm [ (L_{i}^{2} - R_{i}^{2})^{2} {\tilde{m}}^{4} + 4 {A_{i}}^{2} {m}^{2} 
  {\tilde{m}}^{2} ]^{1/2} \}/2, \ \ i = 1, 2 .
\end{eqnarray} 

\noindent{2. Mixing of charged gauginos and higginos:}
    
Defining 
%eq13 
\begin{eqnarray}
  {\Psi}_{j}^{+} & = & ( -i {\lambda}_{L}^{+}, \  (v_{1} {\tilde{\Phi}
  }_{1}^{1} + v {\tilde{H}}_{12})/{(v^{2} + v_{1}^{2})^{1/2}}, \  -i {
  \lambda}_{R}^{+}, \  (v_{2} {\tilde{\Phi}}_{2}^{1} - v {\tilde{H}
  }_{12})/{(v^{2} + v_{2}^{2})^{1/2}} ), \nonumber \\ 
  {\Psi}_{j}^{-} & = & ( -i {\lambda}_{L}^{-}, \  (v_{1} {\tilde{\Phi
  }}_{1}^{\prime 2} + v {\tilde{H}}_{12})/{(v^{2} + v_{1}^{2})^{1/2}}, \  -i {
  \lambda}_{R}^{-}, \  (v_{2} {\tilde{\Phi}}_{2}^{\prime 2} - v {\tilde{H}
  }_{21})/{(v^{2} + v_{2}^{2})^{1/2}} ), \nonumber \\ 
  j & = & 1, 2, 3, 4,  
\end{eqnarray}
the mass eigenstates and their masses are given by 
%eq14 
\begin{eqnarray}
  {\tilde{\chi}}_{i} & = & \left( \begin{array}{c} {\chi}_{i}^{+} \\ {\bar{ 
  \chi}}_{i}^{-}  \end{array} \right), \ \ \ \ \  
  {\chi}_{i}^{+} = V_{ij} {\Psi}_{j}^{+},\ \ \ {\chi}_{i}^{-} = U_{ij} {
  \Psi}_{j}^{-}; \nonumber \\    
  {\tilde{M}}_{1,2} & = & \{ [M_{0}^{2} + 2 g_{L}^{2} ( v^{2} +
  v_{1}^{2} )]^{1/2} \pm M_{0} \} /2, \nonumber \\ 
  {\tilde{M}}_{3,4} & = & \{ [M_{0}^{2} + 2 g_{R}^{2} ( v^{2} 
  + v_{2}^{2} )]^{1/2} \pm M_{0} \} /2 ,
\end{eqnarray} 
where the unitary matrices $U$, $V$ are given by 
%eq15 
\begin{eqnarray}
  U & = & \left( \begin{array}{cc} O_{1} \ & \ 0 \\  0 \ & \ O_{2} \end{array} 
  \right),  \ \ \  V = \left( \begin{array}{cc} {\sigma}_{3} O_{1} \ & \ 0 \\ 
  0 \ & \ {\sigma}_{3} O_{2}  \end{array} \right), \ \ \  {\rm with} \ \ \  
  O_{i} = \left( \begin{array}{cc} cos {\phi}_{i} \ & \ sin {\phi}_{i}  \\ 
  - sin {\phi}_{i} \ & \ cos {\phi}_{i}  \end{array}  \right), \nonumber \\ 
  i & = & 1, 2, \ \ \ cos {{\phi}'_{1}} = [ \tilde{M_{1}} / ( \tilde{M_{1}} +
  \tilde{M_{2}} ) ]^{1/2}, \ \ \  cos {{\phi}'_{2}} = [\tilde{M_{3}} / ( 
  \tilde{M_{3}} + \tilde{M_{4}} ) ]^{1/2} . 
\end{eqnarray} 

\noindent{3. Mixing of neutral gauginos and higginos:}

    Defining 
%eq16  
\begin{eqnarray}
  {\Psi}_{j}^{0} &   & = ( - i {\lambda}_{A}, \  - i {\lambda}_{Z_{1}},\ 
  - i {\lambda}_{2},\  - i {\lambda}_{Z_{2}}, \ - i {\lambda}_{3}  ) ,
\end{eqnarray} 
where $- i {\lambda}_{Z_{2, 3}}$ are given in eq.5, then the mass eigenstates 
and their masses are given by  
%eq17 
\begin{eqnarray}
  &   & {\tilde{\chi}}_{i}^{0} = \left( \begin{array}{c} {\chi}_{i}^{0} \\ 
  {\bar{\chi}}_{i}^{0}  \end{array} \right), \ \ \ {\chi}_{i}^{0} = N_{ij} {
  \Psi}_{j}^{0} , \ \ \ i = 1, \cdot \cdot \cdot , 5; \nonumber \\  
  &   & {\tilde{N}}_{1} = M_{0}, \ {\tilde{N}}_{2,3} = ({m_{Z_{1}}}^{2} + 
  {M_{0}}^{2} /4 )^{1/2} \pm {M_{0}}/2 , \nonumber \\ 
  &   & {\tilde{N}}_{4,5} = ({m_{Z_{2}}}^{2} + {M_{0}}^{2} /4 )^{1/2} \pm 
  {M_{0}}/2 ,
\end{eqnarray}
with the matrix $N$ as   
%eq18  
\begin{eqnarray}
  N & = &  \left( \begin{array}{ccccc} 1 & 0 & 0 & 0 & 0 \\ 0 & cos {\phi}_{1}
  & sin {\phi}_{1} & 0 & 0 \\ 0 & -i sin {\phi}_{1}\ & \ i cos {\phi}_{1} & 0 &
  0 \\ 0 & 0 & 0 & cos {\phi}_{2}  & sin {\phi}_{2} \\ 0 & 0 & 0 & -i sin
  {\phi}_{2} \ & \ i cos {\phi}_{2} \end{array}   \right) \nonumber \\ 
  cos {\phi}_{1} & = & (\frac{{\tilde{N}}_{2}}{{\tilde{N}}_{2} + {\tilde{N}
  }_{3}})^{1/2}, \ \  cos {\phi}_{2} = (\frac{{\tilde{N}}_{4}}{
  {\tilde{N}}_{4} + {\tilde{N}}_{5}})^{1/2} .
\end{eqnarray} 

\noindent{$Mass$ $Difference$ $within$ $an$ $Isodoublet$ $in$ $the$ 
$Soft-Broken$
$Supersymmetric$ $SU(2)_{L} \times SU(2)_{R} \times U(1)_{Y}$ $Model - $}
The SUSY breaking will lead to a different interaction Lagrangian from the one 
given in eq.6, and we should also expect different mass difference within an 
isotopic multiplet from the one given in eq.7. Just for simplicity, instead of 
doing a general numerical studies, I will do an analytic study for one set of 
parameters to show the possibility of obtaining the right sign for the 
mass difference within an isodoublet. 

    Supposed I set ${\mu}_{1} = {\mu}_{2} = {\mu}_{3} = 0$, $M_{1} = M_{2} = 
M_{3} = M_{0}$, $v'_{1} = v^{\prime \prime}_{1} = v_{1}$, $v'_{2} = v^{\prime \prime}
_{2} = v_{2}$ for 
the parameters appearing in the soft-SUSY breaking terms, then I can 
substitute eqs.12, 14, 17 into eq.6, and obtain the Lagrangian for the SUSY 
breaking $SU(2)_{L} \times SU(2)_{R} \times U(1)_{Y}$ model. 
It is not hard to realize that there are two kinds of diagrams 
contributing to ${\Delta} m {\mid}_{d-u}$ from the SUSY breaking Lagrangian. 

    Kind 1 is of the form $P_{L} \ {\Gamma} \ P_{R} \  {\rm or} \  P_{R} \ {
\Gamma} \ P_{L}$, which, being similar to the integrals I got in Section II, 
would be proportional to the fermion mass of the isodoublet.

    Kind 2 is of the form $P_{L} \ {\Gamma} \ P_{L} \  {\rm or} \  P_{R} \ {
\Gamma} \ P_{R}$, which we did not see before. As we will see, this will play 
a very important role in getting a right sign for ${\Delta} m {\mid}_{d-u}$.

    Define  
%eq19  
\begin{eqnarray} 
  J (m, m_{1}, m_{2}, m_{3}) & = & \frac{-i m^{2}}{16 {\pi}^{2}} {\int}_{0
  }^{1} dz{\ } ln{\frac{z(z-1) m^{2} + z {m_{3}}^{2} + (1-z) m_{1}^{2}}{z(z-1) 
  m^{2} + z m_{3}^{2} + (1-z) m_{2}^{2}}} .
\end{eqnarray} 
If I use ${\Delta} m_{1,2} |_{d-u}$ to represent contributions from kind 1, 2 
respectively, and further set ${\mu}_{1I} = {\mu}_{2I}$, 
${\mu}_{1II} = {\mu}_{2II}$, which also lead to ${\theta}_{1} = {\theta}_{2}$, 
detailed calculations show that   
%eq20  
\begin{eqnarray}
  {\Delta} m_{1} |_{d-u} & > & - \frac{\alpha m}{2 \pi} {\int}_{0}^{1} dx
  (1+x)[ d_{1} ln(1+ \frac{1-x}{x^{2}} \frac{m_{Z_{1}}^{2}}{m^{2}}) + d_{2}
  ln(1 + \frac{1-x}{x^{2}} \frac{m_{Z_{2}}^{2}}{m^{2}}) ] \nonumber \\ 
  {\Delta} m_{2} |_{d-u} & = & ({i {\tilde{N}}_{m}} / 2 m^{2} ) Re ( N_{m,j+ 
  c_{j}} N_{m,k+c_{k}}) sin 2 {\theta}_{1} J (m, {\mu}_{1I}, {\mu}_{1II}, 
  {\tilde{N}}_{m}) g_{Y} y_{L} \cdot \nonumber \\ 
  &   & \ \ \ \ \ \ \ \ (g_{L} a_{j1} a_{k3} + g_{R}  a_{j3} a_{k2} ) ,
\end{eqnarray} 
where in ${\Delta} m_{1} |_{d-u}$, some terms give positive, and some negative
contributions to ${\Delta} m |_{d-u}$, and in ${\Delta} m_{2} |_{d-u}$, 
$c_{1} = c_{2} = 0$, $c_{3} = 1$; $j, k = 1, 2, 3$. Since 
%eq21  
\begin{eqnarray}
  J (m, {\mu}_{1I}, {\mu}_{1II}, {\tilde{N}}_{m}) > J (m, {\mu}_{1I}, 
  {\mu}_{1II}, {\tilde{N}}_{n}) \ \ \ {\rm for} \ {\mu}_{1I} > {\mu}_{1II}, 
  \ {\tilde{N}}_{m} < {\tilde{N}}_{n} ,
\end{eqnarray} 
using eqs.21, 18, 17, I can further rewrite ${\Delta} m_{2} |_{d-u}$ in eq.20 
as 
%eq22  
\begin{eqnarray}
  {\Delta} m_{2} |_{d-u} & > & ( {i g_{L} e^{2} M_{0} sin 2 {\theta}_{1}} 
  / m^{2} ) \{ d_{1} [ J (m, {\mu}_{1I}, {\mu}_{1II}, M_{0}) - J  (m, {
  \mu}_{1I}, {\mu}_{1II}, {\tilde{N}}_{3})] + \nonumber \\  
  &   & d_{2} [J (m, {\mu}_{1I}, {\mu}_{1II}, M_{0}) - J (m, {\mu}_{1I}, 
  {\mu}_{1II},{\tilde{N}}_{5})] \}  \nonumber \\
  & = & {\kappa} M_{0} > 0 ,
\end{eqnarray}
provided we have $M_{0} < {\tilde{N}}_{3}$ and $M_{0} < {\tilde{N}}_{5}$, 
i.e. $m_{Z_{1}} > {\sqrt{2}} M_{0}$ and $m_{Z_{2}} > {\sqrt{2}} M_{0}$. 

    For $m_{Z_{1}}$ and $m_{Z_{2}}$ are at least in the order of magnitude of
$10^2\ Gev$, it is not hard to have $M_{0} >> m$ ($\sim$ a few $Mev$) but still 
satisfy the requirement $m_{Z_{1}} > {\sqrt{2}} M_{0}$ and $m_{Z_{2}} > {\sqrt{
2}} M_{0}$. This condition is not inconsistent with the common expectation in 
supersymmetry phenomenology[4]. 

    We notice from eq.20 that ${\Delta} m_{1} |_{d-u}$ at most increases with 
logarithm of $m_{Z_{i}} /m$, but detailed analysis to eqs.22, 19 shows that 
$\kappa$ increases with ${\mu}_{1I} / {\mu}_{1II}$, i.e. $\kappa$ can be a 
not very small value by proper choices of ${\mu}_{1I}$ and ${\mu}_{1II}$, e.g. 
$\kappa \sim 0.001$[5], and also $M_{0} >> m$
($\sim$ few $Mev$). Therefore, I can be definite to expect that 
%eq23 
\begin{eqnarray}
  {\Delta} m|_{d-u} = {\Delta} m_{1} |_{d-u} + {\Delta} m_{2} |_{d-u} 
  \sim {\Delta} m_{2} |_{d-u} > 0 ,
\end{eqnarray} 
for proper choices of ${\mu}_{1I}$ and ${\mu}_{1II}$ in the case of  
${\mu}_{1} = {\mu}_{2} = {\mu}_{3} = 0$, $M_{1} = M_{2} = M_{3} = M_{0}$, 
$v'_{1} = v^{\prime \prime}_{1} = v_{1}$, $v'_{2} = v^{\prime \prime}_{2} = v_{2}$, and 
${\mu}_{1I} = {\mu}_{2I}$, ${\mu}_{1II} = {\mu}_{2II}$.  

\noindent{$Discussion -$} In this papaer, the possibility of obtaining the
right sign 
for the mass difference within an isotopic multiplets in a supersymmetric gauge 
theory is raised. In the scheme I used, SUSY breaking, which is what we 
expected if supersymmetry is really a symmetry in the real world, play a 
very important role. The model is not inconsistent with the Standard Model and 
the present experimental limits[4].  Certainly, I only choose 
a special set of the parameters for calculation simplification. A thorough 
numerical study for the parameters would be nice for obtaining the particle 
spectrums. If the mechanism I used really say something about the nature, we 
may be able to use ${\Delta} m|_{d-u}$ as a constraint to the parameters 
appearing in soft-supersymmetry breaking. Of course, I only raise a 
possibility here. Even under the current framework, much more works are still 
needed, especially considering the ideas developed in the last few decades in 
field theories and particle physics.

This work is supported by the DOE through Grant No. UCR (DE-FG03-86ER40271).

%
%TABLE
%TABLE 1
\begin{table}
\caption{The fields in the SUSY $SU(2)_{L} \times SU(2)_{R} \times U(1)_{Y}$
model. Note: (1) The charge is obtained via $Q = T_{L3} + T_{R3} + Y/2$; (2)
${\psi}_{i} = \bigg({{{\psi}_{iL}}\atop {{\bar{\psi}}_{iR}}}\bigg) $; (3) For
$\bigg({p\atop n}\bigg)$ and $\bigg({u\atop d}\bigg)$, we will have
$y_{L} = - y_{R} = 1$, and $y_{L} = - y_{R} = 1/3 $ respectively.}

\vspace{3.0mm}

\begin{tabular}{|l|l|l|l|l|r|}      \hline
{} & Boson Fields & Fermionic Partners & $SU(2)_L$ & $SU(2)_R$ & $Y$ \\ \hline
Gauge multiplets & ${\vec{A}}_{\mu}^{L}$ & ${\vec{\lambda}}_{L}$ & Triplet
& {} & 0 \\
{} & ${\vec{A}}_{\mu}^{R}$ & ${\vec{\lambda}}_{R}$ & {} & Triplet & 0 \\
{} & $B_{\mu}$ & ${\lambda}'$ & Singlet & Singlet & 0 \\     \hline
Matter multiplets & {} & {} & {} & {} & {} \\
Scalar fermions & ${\tilde{\psi}}_{jL} = ( {\tilde{\psi}}_{1L},\
{\tilde{\psi}}_{2L} )$ & ${\psi}_{jL} = ( {\psi}_{1L},\ {\psi}_{2L} )$ &
Doublet & Singlet & $y_{L}$ \\
{} & ${\tilde{\psi}}_{jR} = (-{\tilde{\psi}}_{2R}^{*},\ {\tilde{\psi}}_{
1R}^{*})$ & ${\psi}_{jR} = (-{\psi}_{2R},\ {\psi}_{1R})$ & Singlet &
Doublet  & $y_{R}$ \\
Higgs Bosons & ${\Phi}_1 = ({\Phi}_1^{1},\ {\Phi}_1^{2})$ & ${\tilde{\Phi}}_1
= ({\tilde{\Phi}
}_1^{1},\ {\tilde{\Phi}}_1^{2})$ & Doublet & Singlet & 1 \\
{} & ${\Phi}'_1 = ({\Phi}_1^{\prime 1},\ {\Phi}_1^{\prime 2})$ & ${\tilde{\Phi}}'_1 =
({\tilde{\Phi}}_1^{\prime 1},\
{\tilde{\Phi}}_1^{\prime 2})$ & Doublet & Singlet & -1 \\
{} & ${\Phi}_2 = ({\Phi}_2^{1},\ {\Phi}_2^{2})$ & ${\tilde{\Phi}}_2 =
({\tilde{\Phi}}_2^{1}, \ {\tilde{\Phi}}_2^{2})$ & Singlet & Doublet & 1 \\
{} & ${\Phi}'_2 = ({\Phi}_2^{\prime 1},\ {\Phi}_2^{\prime 2})$ & ${\tilde{\Phi}}'_2 =
({\tilde{\Phi}}_2^{\prime 1},\ {\tilde{\Phi}}_2^{\prime 2})$ & Singlet & Doublet & -1 \\
{} & $H= \left( \begin{array}{cc}  h_{11} {\ } & {\ } h_{12} \\
h_{21} {\ } & {\ } h_{22}  \end{array}   \right)$ & $\tilde{H}= \left(
\begin{array}{cc}  {\tilde{h}}_{11} {\ } & {\ } {\tilde{h}}_{12} \\
{\tilde{h}}_{21} {\ } & {\ } {\tilde{h}}_{22}  \end{array}   \right)$ &
Doublet & Doublet & 0 \\
{} & $N_1$, $N_2$, $N_3$ & ${\tilde{N}}_1$, ${\tilde{N}}_2$, ${\tilde{N}}_3$ &
Singlet & Singlet & 0 \\     \hline
\end{tabular}
\end{table}

\vspace{7mm}

\noindent{$\ast$ Correspondence address: P25, MS H846,
Los Alamos National Laboratory, Los Alamos,} 

\noindent{\,\ \ NM 87545.}

\end{document}